\documentclass[11pt,letterpaper]{article}

\usepackage{amsmath,amstext,amsthm,amsfonts}
\usepackage[dvips]{graphicx}
\usepackage{amssymb}
\usepackage{latexsym}
\usepackage[hmargin=35mm,top=35mm,bottom=35mm]{geometry}
\usepackage{hyperref}

\theoremstyle{plain}
\theoremstyle{definition}
\newtheorem{theorem}{Theorem}
\newtheorem{conjecture}{Conjecture}
\newtheorem{problem}{Problem}

\hypersetup{colorlinks=true,urlcolor=blue,linkcolor=black,citecolor=black}

\begin{document}

\title{Activated Random Walkers:\\ Facts, Conjectures and Challenges}
\author{Ronald Dickman, Leonardo T. Rolla, Vladas Sidoravicius}
\date{December 29, 2009}
\maketitle

\begin{abstract}
We study a particle system with hopping (random walk) dynamics on the integer
lattice $\mathbb Z^d$.  The particles can exist in two states, active or inactive
(sleeping); only the former can hop. The dynamics conserves the number of
particles; there is no limit on the number of particles at a given site. Isolated active particles fall asleep at
rate $\lambda > 0$, and then remain asleep until joined
by another particle at the same site.
The state in which all particles are inactive is absorbing.
Whether activity continues at long times depends on the relation
between the particle density $\zeta$ and the sleeping rate $\lambda$.
We discuss the general case, and then, for the one-dimensional totally asymmetric
case,
study the
phase transition between an active phase
(for sufficiently large particle densities and/or small $\lambda$) and an absorbing one.
We also present arguments regarding the asymptotic mean hopping velocity in the active phase, the rate of fixation in the absorbing phase, and survival
of the infinite system at criticality.  Using mean-field theory and Monte Carlo
 simulation, we locate the phase boundary. The phase transition appears to be
continuous in both the symmetric and asymmetric versions of the process, but
the critical behavior is very different.  The former case is characterized by
simple integer or rational values for critical exponents ($\beta = 1$, for
example), and the phase diagram is in accord with the prediction of mean-field
theory.  We present evidence that the symmetric version belongs to the
universality class of conserved stochastic sandpiles, also known as
conserved directed percolation. Simulations also reveal
an interesting transient phenomenon of damped oscillations in the activity
density.
\end{abstract}


This preprint has the same numbering of sections, equations, figures, problems and theorems as ``\emph{J. Stat. Phys. 138 (2010), 126-142.}''
The the published article is available as Open Access at
\href
{\detokenize{http://dx.doi.org/10.1007/s10955-009-9918-7}}{doi:10.1007/s10955-009-9918-7}.

\section{Introduction}
\label{sec1}

Interacting particle systems with conservation have attracted great interest in
physics, probability, and allied fields, in part because they afford simple
examples of phase transitions in systems maintained far from equilibrium. In
these models the local dynamics conserves the number of particles, although
certain sites may act as particle sources or absorbers. One broad important
class of models subsumes exclusion models, in which particles interacting via
on-site exclusion (and possibly an additional short-range interaction) execute
biased hopping on a lattice.
Important examples are driven diffusive systems~\cite{katz-lebowitz-spohn-83,katz-lebowitz-spohn-84,schmittmann-zia-95,marro-dickman-99} and the totally asymmetric exclusion process~\cite{schutz-01,krug-91,derrida-evans-hakim-pasquier-93}.
In another class of models there is no
exclusion (any number of particles may occupy the same site) but the particles
exist in two states that may be termed active and inactive, such that
activation of an inactive particle requires the intervention of one or more active ones.
This class includes so-called conserved lattice gases~\cite{rossi-pastorsatorras-vespignani-00,lubeck-02,lubeck-02-1,lubeck-heger-03,lubeck-heger-03-1} and stochastic sandpile models~\cite{manna-90,manna-91,dickman-alava-munoz-peltola-vespignani-zapperi-01,dickman-tome-oliveira-02,dickman-06}.
Such models exhibit self-organized criticality~\cite{bak-tang-wiesenfeld-87,bak-tang-wiesenfeld-88,grinstein-95,dhar-99} when coupled to a suitable control mechanism~\cite{dickman-munoz-vespignani-zapperi-00,dickman-02}.

In this paper we study a system of activated random walkers (ARW) on the
lattice. For theoretical analysis, it is convenient to define the model (ARW1)
on the infinite integer lattice $\mathbb Z^d$. We assume, in this case, that there
are infinitely many particles in the system, each of which can be in one of two
states: $A$ (active) or $S$ (inactive or sleeping). Each $A$-particle performs
an independent, continuous time, simple symmetric random walk on $\mathbb Z^d$,
with the same jump rate, which we assume, without loss of generality, to be
equal to $1$. When an $A$-particle jumps to a site with an $S$-particle or
particles, any such particle at this site is immediately activated (i.e.,
switches to state $A$). Each isolated $A$-particle goes to sleep (switches to
state $S$), at a rate $\lambda
>0$.  [From this rule it follows that if two or more particles
occupy the same site, then they are all of type $A$ or all of type $S$ (the
latter situation can only arise in the initial condition).] Since $S$-particles
are immobile, at any given site, at most one $A$-particle can go to sleep, and
(if undisturbed), remain in state $S$ forever after.
The limit $\lambda \to \infty$ corresponds to the model studied by Jain \cite{jain-05}, in which any isolated particle immediately becomes immobile.

We assume that initially the particles are distributed according to a product Poisson measure with mean $\zeta$, and all the particles are active.

In numerical studies the following one-dimensional model (ARW2) is used.
The system is a chain of $L$ sites with either periodic or open boundaries.
Initially $N$ particles are randomly placed in the system. (In the case of
periodic boundaries the particle number is conserved.)  As in ARW1, each
nonisolated $A$-particle hops at unit rate. An isolated $A$-particle has a somewhat
different dynamics: it hops at rate $p \leq 1$ and goes to sleep at rate $q =
1-p$. Thus in ARW2 isolated $A$-particles have a smaller hopping rate than
nonisolated ones, while in the ARW1 all $A$-particles have the same jump rate.
While certain details such as the phase boundary may differ between the two
versions, we expect the global properties to be the same.
For the question of whether or not the system fixates, the two models are equivalent via $\lambda = q/p$.
The case $p=0$ again represents the model studied in \cite{jain-05}.

A natural generalization of the model is to allow the $A$-particles to execute
a {\it biased} random walk.
We shall in fact be particularly interested in the completely asymmetric case.
Another possible generalization is to assume
that each $S$-particle is activated at rate $0 < \alpha \leq + \infty$, when it
shares its site with an $A$-particle. One could go even further, by assuming
that the rate at which an $A$-particle activates $S$ particles at a given site
depends on the number of $S$ particles at the site (zero-range rule). There is
however a substantial difference between the case $\alpha < + \infty$, and the
case $\alpha = + \infty$, described at the beginning of this section.
If $\alpha < + \infty$, during evolution the $A$-particles may share a site along with one or more $S$-particles, so that this model is a kind of dynamic contact process.
If $\alpha = + \infty$,
the situation changes, and we believe this model belongs to the universality
class of stochastic conserved sandpiles. In this work, we only consider the
case $\alpha = + \infty$ (instantaneous reactivation).

The primary motivation for the present study is the stochastic conserved
sandpile, generally known as Manna's model~\cite{manna-90,manna-91}.
In infinite volume,
this model is defined as follows. We assume that initially there are infinitely
many particles, distributed in such a way that at each site of $\mathbb Z^d$ we
have Poisson mean $\zeta>0$ number of particles. Each site is equipped with an
exponential rate $1$ clock, and each time a clock rings at a site bearing $2d$
or more particles, $2d$ particles move from the site to randomly chosen nearest
neighbors.  That is, (differently from the deterministic Bak-Tang-Wiesenfeld
sandpile model~\cite{bak-tang-wiesenfeld-87,bak-tang-wiesenfeld-88}), each particle chooses its direction among the $2d$
possibilities with probability $(2d)^{-1}$, independent of any other particle.
In contrast to the deterministic sandpile~\cite{dhar-99}, very little is known rigorously about this system, and the ARW model is a reasonable caricature that seems to capture some essential aspects of Manna's model.

The ARW model may also be viewed as a special case of a diffusive epidemic
process, in which an infected particle performs a simple symmetric random walk
with jump rate $D_B$, and recuperates at a given rate, while a healthy
particle performs a simple symmetric random walk with jump rate $D_A$.
(Healthy particles are infected on contact with infected ones.)
The ARW model corresponds to $D_A = 0$. The generalized CP (in the case $D_A = D_B$) was proposed in the late 1970's by Spitzer, and later was studied in detail in~\cite{kesten-sidoravicius-03,kesten-sidoravicius-05,kesten-sidoravicius-06,kesten-sidoravicius-08}.
The diffusive epidemic process has also been studied via renormalization group and numerical simulation~\cite{kree-schaub-schmittmann-89,wijland-oerding-hilhorst-98,oerding-wijland-leroy-hilhorst-00,freitas-lucena-silva-hilhorst-00-1,fulco-messias-lyra-01,fulco-messias-lyra-01-1,janssen-01,dickman-souzamaia-08,souzamaia-dickman-07}.
A general conclusion from these
studies is that there are three distinct regimes of critical behavior, for
$D_A < D_B$, $D_A = D_B$ and $D_A > D_B$.  It is not yet
clear whether the ARW model falls in the first regime, or, alternatively, that
$D_A =0$ marks a special case.

The ARW with symmetric hopping is closely related to the conserved lattice gas model (CLG)~\cite{rossi-pastorsatorras-vespignani-00,lubeck-02,lubeck-02-1,lubeck-heger-03,lubeck-heger-03-1}, the principal difference being that in the CLG a site can be occupied by at most one particle, while active particles are those having at least one occupied neighbor.
The CLG and the conserved stochastic sandpile share the same essential features, namely, a continuous phase transition between an active and an absorbing state, conservation of particles, and coupling between the order parameter (activity) and particle density, with the particle configuration frozen in regions devoid of activity.
There is evidence~\cite{ramasco-munoz-santos-04,dornic-chate-munoz-05,dickman-06} that the CLG and conserved stochastic sandpile exhibit the same critical behavior, and we should expect the same to apply to the symmetric ARW model.

Numerical analysis and some general theoretical arguments suggest that the ARW
model exhibits a phase transition in the parameters $\lambda$ and $\zeta$, and
that there should be two distinct regimes:

{\bf i)} Low particle density. There is a phase transition in $\lambda$ in this
case, namely if $\lambda$ is large enough, then system locally fixates, i.e.
for any finite volume $\Lambda$ there is almost surely a finite time
$t_{\Lambda}$ such that after this time there are no $A$ particles within
$\Lambda$. If $\lambda$ is small enough there is no fixation, and we expect
that there is a limiting density of active particles in the long-time limit.

{\bf ii)} High particle density. In this case there is no phase transition.
For any $\lambda >0$, the system does not fixate.

The balance of this paper is organized as follows.
In Section~\ref{sec2mft} we study ARW2 with totally asymmetric walks via mean-field theory.
Section~\ref{sec3simulation} contains a detailed study of the critical behavior for totally asymmetric walks, including power laws and scaling relations, via Monte Carlo simulations, and ends mentioning the symmetric case, which belongs to another universality class.
In Section~\ref{sec4resultschalenges} we quote the few known mathematical results and discuss a collection of open problems.

\section{Mean-field theory}
\label{sec2mft}

In this section we develop a mean-field theory (MFT) for the ARW2
model defined in Section~\ref{sec1}.  As is usual in this type of approach, we treat
the state of each site as statistically
independent.
Although the discussion is formulated for the totally asymmetric case, this `one-site' approximation in fact yields the same predictions for the symmetric version.

For $n \geq 1$, let $p_n (t)$ be the fraction of sites having exactly $n$ $A$-particles.
We denote the fraction of sites occupied by an $S$-particle by
$p_1' (t)$, and the fraction of vacant sites by $p_0 (t)$. Normalization
implies,
\begin{equation*}
p_1' + \sum_{n=0}^\infty p_n = 1
\end{equation*}
while the particle density is
\begin{equation*}
\zeta = p_1' + \sum_{n=0}^\infty n p_n \;,
\end{equation*}
Let
\begin{equation*}
\rho^* = \sum_{n=2}^\infty n p_n ,
\end{equation*}
so that the density of $A$-particles is $\rho_a = \rho^* + p_1 =
\zeta - p_1'$.

We now obtain the equations of motion for the $p_n$, starting
with $n=0$.  The rate of transitions into $n=0$ is $p p_1$, that
is, to enter the state 0 a site must have a single $A$-particle,
which leaves at rate $p$.  The rate of transitions out of state
0 is:
\[
\sum_{n=2}^\infty n p_{n,0} + p p_{1,0}
\]
where $p_{n,m}$ is the joint probability for a pair of nearest-neighbor sites
$j$ and $j+1$ to harbor $n$ and $m$ $A$-particles, respectively. ($p_{1,0}$ is
the joint probability for a site to be empty, and its neighbor on the left
occupied by a single $A$-particle.)  The reason is that to exit state zero, a
site must be in that state and have a nearest neighbor on the left with one or
more particles capable of jumping onto it.

The {\it mean-field approximation} consists in factoring all joint
probabilities: $p_{n,m} \to p_n p_m$.   Combining the rates
for transitions into and out of state zero, and applying the
mean-field factorization, we obtain:

\begin{equation*}
\frac{dp_0}{dt}  = p p_1 - \tilde{\rho} p_0 ,
\end{equation*}
where \[\tilde{\rho} \equiv \rho^* + p p_1.\] Proceeding in the same
manner we obtain equations of motion for the other one-site
probabilities:
\begin{equation*}
\frac{dp_1}{dt}  = \tilde{\rho}( p_0 - p_1) + 2 p_2 - p_1
\end{equation*}
\begin{equation*}
\frac{dp_1'}{dt}  = q p_1 - \tilde{\rho} p_1'
\end{equation*}
\begin{equation*}
\frac{dp_2}{dt}  = \tilde{\rho}( p_1 + p_1' - p_2) - 2p_2 + 3 p_3
\end{equation*}
and for $n \geq 3$,
\begin{equation}
\label{gen}
\frac{dp_n}{dt}  = \tilde{\rho}( p_{n-1}  - p_n) - np_n + (n+1) p_{n+1}.
\end{equation}
These equations conserve normalization and the particle density $\zeta$.
For $\zeta < 1$ there is an inactive solution, $p_1' = \zeta$, $p_0 = 1-\zeta$.

We seek an active stationary solution by introducing a `quasi-Poisson' ansatz,

\begin{equation*}
p_n = {\cal A } \frac{\lambda^n}{n!} \;\;\;\; \mbox{for} \;\; n \geq 3.
\label{qpoisson}
\end{equation*}

\begin{figure}[!htb]
\centering
\includegraphics[width=10cm]{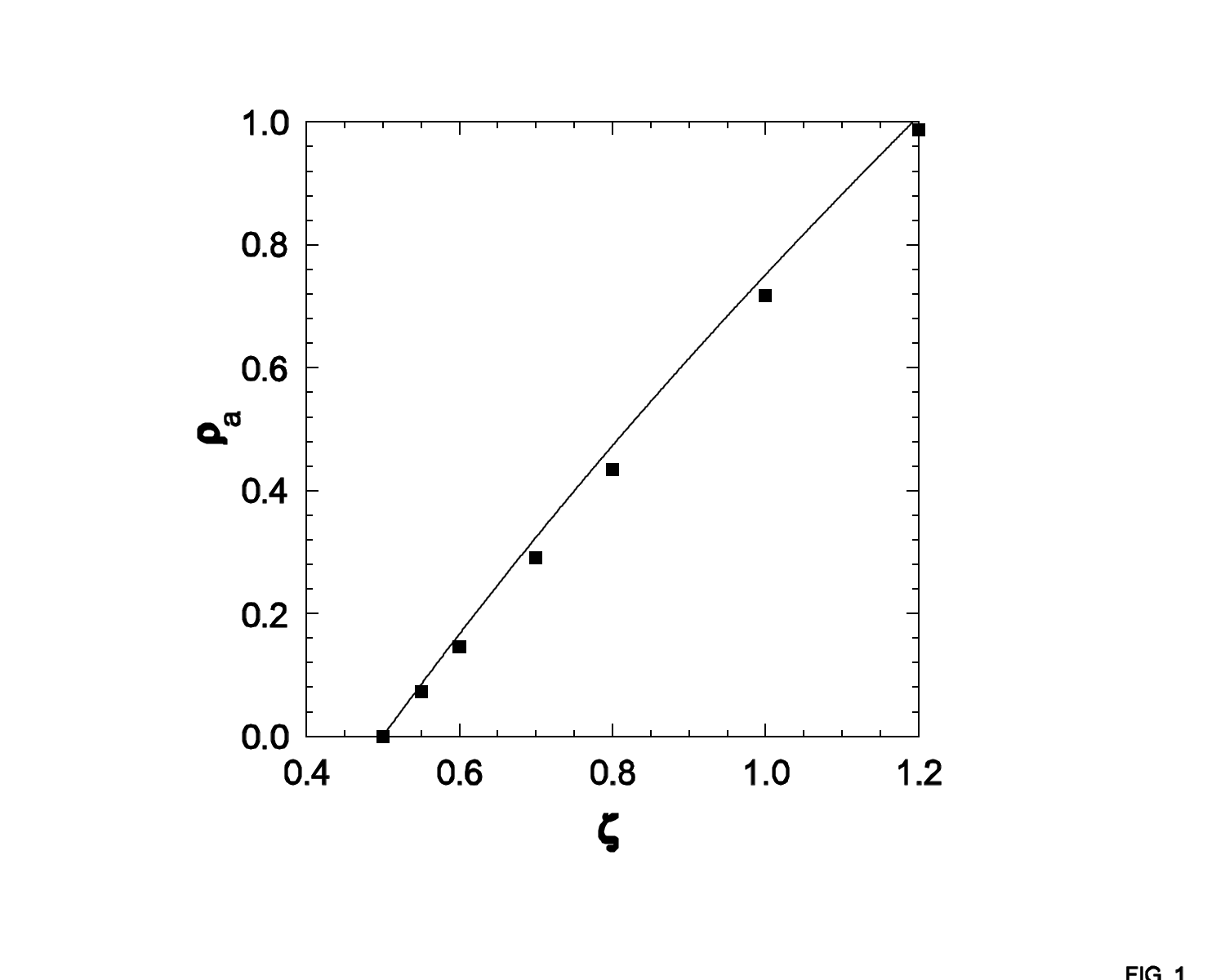}
\caption{Comparison of simulation (squares) and mean-field theory (solid line) for the stationary activity density $\zeta_a$ at $p=0.5$.}
\label{fig1}
\end{figure}

Substituting this hypothesis in Eq.~\eqref{gen} (with $n\geq 4$) one finds
$\lambda = \tilde{\rho}$.
Using the equations for $n=3$, 2, 1, and 0, we obtain $p_2 = {\cal A}\tilde{\rho}^2/2$, $p_1 + p_1' = {\cal A} \tilde{\rho}$, $p_1 = {\cal A}\tilde{\rho}^2/(q + \tilde{\rho})$, and $p_0 = p p_1/\tilde{\rho}$.
Normalization then implies
\begin{equation*}
{\cal A } = \left[ e^{\tilde{\rho}} + \frac{p\tilde{\rho}}{q +
\tilde{\rho}} - 1 \right]^{-1}.
\end{equation*}
With the stationary distribution in hand, we may write
\begin{equation}
\label{den}
\zeta = \rho^* + p_1 + p_1' = {\cal A}(\tilde{\rho}) \tilde{\rho}e^{\tilde{\rho}} = \frac{\tilde{\rho}e^{\tilde{\rho}}} {e^{\tilde{\rho}} + p\tilde{\rho}/(q + \tilde{\rho}) - 1}.
\end{equation}
A plot of the active particle density $\rho_a = \zeta - p_1'$
versus $\zeta$, is shown in Figure~\ref{fig1}, for $p=1/2$.

To locate the critical point $\zeta_c$, we evaluate the limit of the r.h.s.\ of Eq.~\eqref{den} as $\tilde{\rho} \to 0$, yielding
\[\zeta_c = 1-p.\]
For densities smaller than $\zeta_c$,
$\tilde{\rho} = 0$ and $\zeta = p_1'$, i.e., all particles
eventually go to sleep.  The stationary activity density grows
$\propto \zeta - \zeta_c$ for $\zeta > \zeta_c$. Numerical
integration of the mean-field equations shows that the solution
indeed converges to the stationary one found above. For an initial
Poisson distribution, or one in which a fraction $\zeta <1$ of
sites are singly occupied and the rest vacant, the approach to the
stationary state is monotonic, and exponentially rapid away from
the critical point.  At the critical point the activity density
decays algebraically, $\rho_a \sim t^{-1}$.  These are the usual
characteristics of a mean-field theory for an absorbing-state
phase transition~\cite{marro-dickman-99}.

The mean-field analysis is readily extended to the two-site approximation, in
which the dynamical variables are the joint probabilities $p_{i,j}$ for a pair
of neighboring sites to have occupations $i$ and $j$~\cite{marro-dickman-99}. (In this
case three-site probabilities are approximated so: $p_{i,j,k} \simeq
p_{i,j}p_{j,k}/p_j$.)  The pair approximation again yields $\zeta_c = 1-p$. The
stationary density of active sites is slightly less than in the simple MF
approximation (a reduction of about 6\%, near the critical point), and the
relaxation at the critical point again follows $\rho_a \sim t^{-1}$.
Finally, we note that for model ARW1 (jump rate unity for all $A$ particles, sleep rate $\lambda$ for isolated $A$ particles), the mean-field analysis yields the critical density
\[\zeta_c = \lambda /(1+\lambda),\]
in agreement with simulations and rigorous results~(Theorem~\ref{theo1tarw}).

\section{Simulation results}
\label{sec3simulation}

We performed Monte Carlo simulations of the ARW2 model on rings of $L=$ 100 to
8000 sites.  Stationary and time-dependent properties were determined from
averages over 10$^5$ - 10$^6$ independent realizations of the process, starting
from an initial configuration in which all particles are in state $A$. In
simulations, we select an $A$-particle (a list of such particles is
maintained), and if it is not isolated, it jumps to the right. If the selected
particle is isolated then it goes to sleep with probability $q$, and jumps
forward with probability $p=1-q$.  The time increment associated with each
event is $\Delta t = 1/n$, with $n$ the number of $A$ particles just before the
event.
(For $\zeta < 1$, a finite system must eventually become trapped in an absorbing configuration.
In practice, however, the lifetime of the
quasi-stationary metastable state observed in simulations is very long for $p >
p_c$.  It appears, moreover, that the quasi-stationary properties observed in
simulations converge to a well-defined limit as the system size $L \to \infty$,
that is, to the true stationary properties of the infinite system.)

\subsection{Phase diagram}

We studied the stationary density of $A$-particles $\rho_a$, the moment ratio
$m = \langle \rho_a^2 \rangle / \rho_a^2$, and the survival probability
$P_s(t)$, that is, the probability that not all walkers are asleep.  The
variation of the stationary activity density with walker density $\zeta$ (for
fixed sleeping probability $q=1/2$) is shown in Figure~\ref{fig1}.  Note that the data
represent extrapolations to the infinite-size limit based on results for
systems of size 100, 200,...,3200. The simulation result is very close to,
although systematically smaller than, the mean-field prediction.  (The small
but nonzero
difference between simulation and MFT cannot be attributed to a finite size
effect.)   The pair approximation is in somewhat better agreement with
simulation. For example, the stationary activity density near the critical
follows
\begin{equation*}
\rho_a = B (\zeta-\zeta_c).
\end{equation*}
For $p=1/2$, for example, simple MFT yields an amplitude of $B =
1.70$; the amplitude in the pair approximation is 1.60, while simulation yields $B = 1.46$.

Of particular interest is the location of the phase boundary. For
$\zeta \leq 1$ (with $p$ fixed at 1/2), the data for $\rho_a$ fall
very nearly on a straight line that intercepts $\rho_a = 0$ at
$\zeta=0.5$, as predicted by MFT. For any finite system size the
quasi-stationary activity density at $\zeta=1/2$ is nonzero, but
$\rho_a$ approaches zero with increasing system size.
Finite-size scaling theory~\cite{privman-90,fisher-barber-72,fisher-71} predicts a power-law dependence of the stationary order parameter $\rho_a$ on system size along the critical line:
\begin{equation}
\label{rhofss}
\rho_a (p_c,\zeta,L) \sim L^{-\beta/\nu_\perp}
\end{equation}
where $\beta$ and $\nu_\perp$ are the critical exponents
associated, respectively, with the order parameter and the
correlation length~\cite{marro-dickman-99}. (Away from the critical line
$\rho_a$ converges exponentially to its stationary value as $L \to
\infty$.)  A similar picture holds for the lifetime
$\tau(p,\zeta,L)$ defined in terms of the {\it survival
probability} $P_s(t)$.
In a finite system, for $\zeta < 1$, we expect $P_s \sim \exp(-t/\tau)$.
Along the critical line, finite-size scaling theory predicts
\begin{equation}
\label{taufss}
\tau (p_c,\zeta,L) \sim L^{\nu_{||}/\nu_\perp}
\end{equation}
with $\nu_{||}$ the critical exponent associated with the correlation time.
Rather than attempt a systematic justification of these scaling ideas here, we simply note that the behaviors implied by Eqs.~\eqref{rhofss} and~\eqref{taufss} have been amply confirmed in studies of many absorbing-state transitions (as well as in equilibrium critical phenomena), including conserved stochastic sandpiles.

\begin{figure}[!htb]
\centering
\includegraphics[width=10cm]{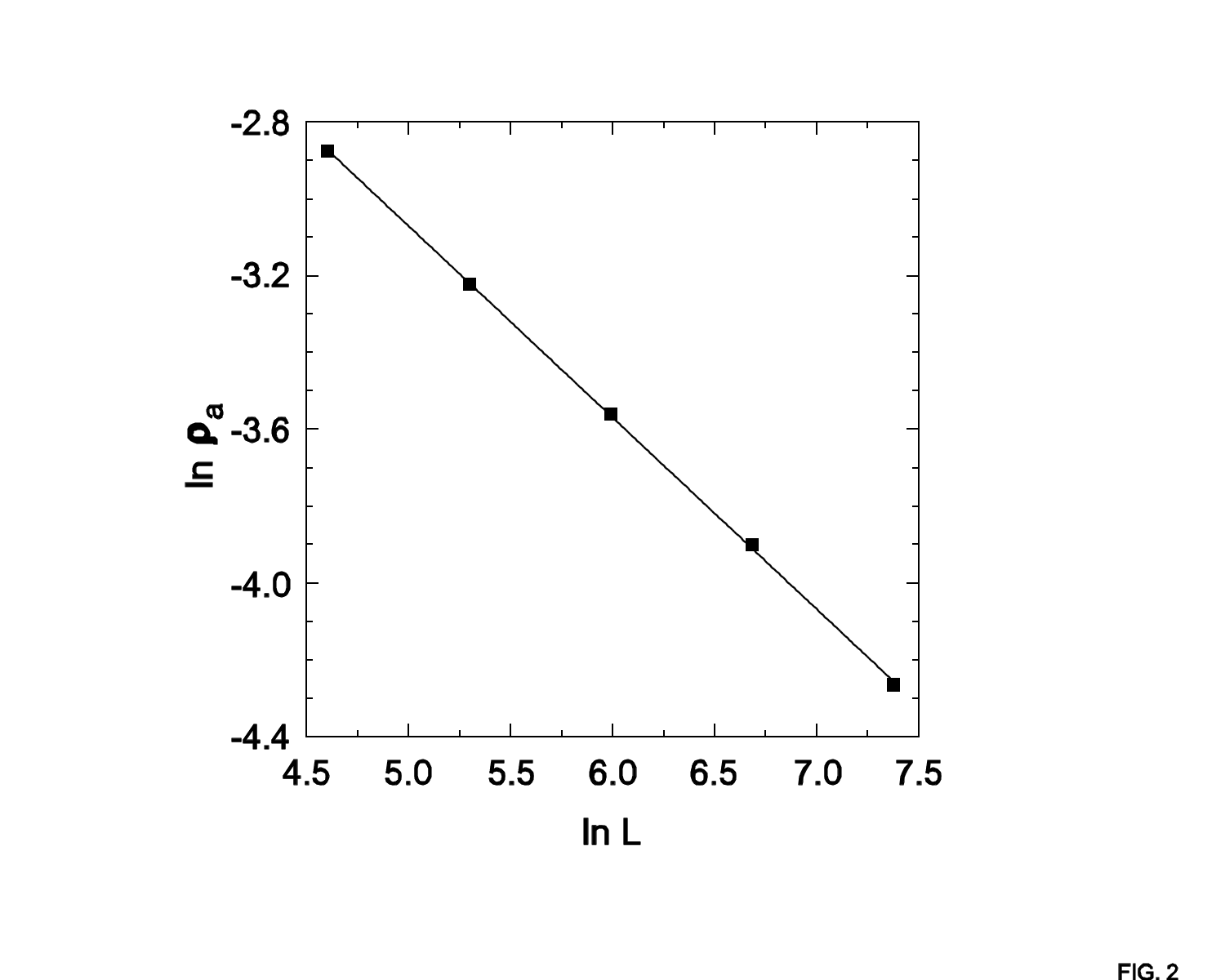}
\caption{%
Stationary order parameter versus system size
for $\zeta=0.25$ and $p=0.75$.
}
\label{fig2}
\end{figure}

\begin{figure}[!htb]
\centering
\includegraphics[width=10cm]{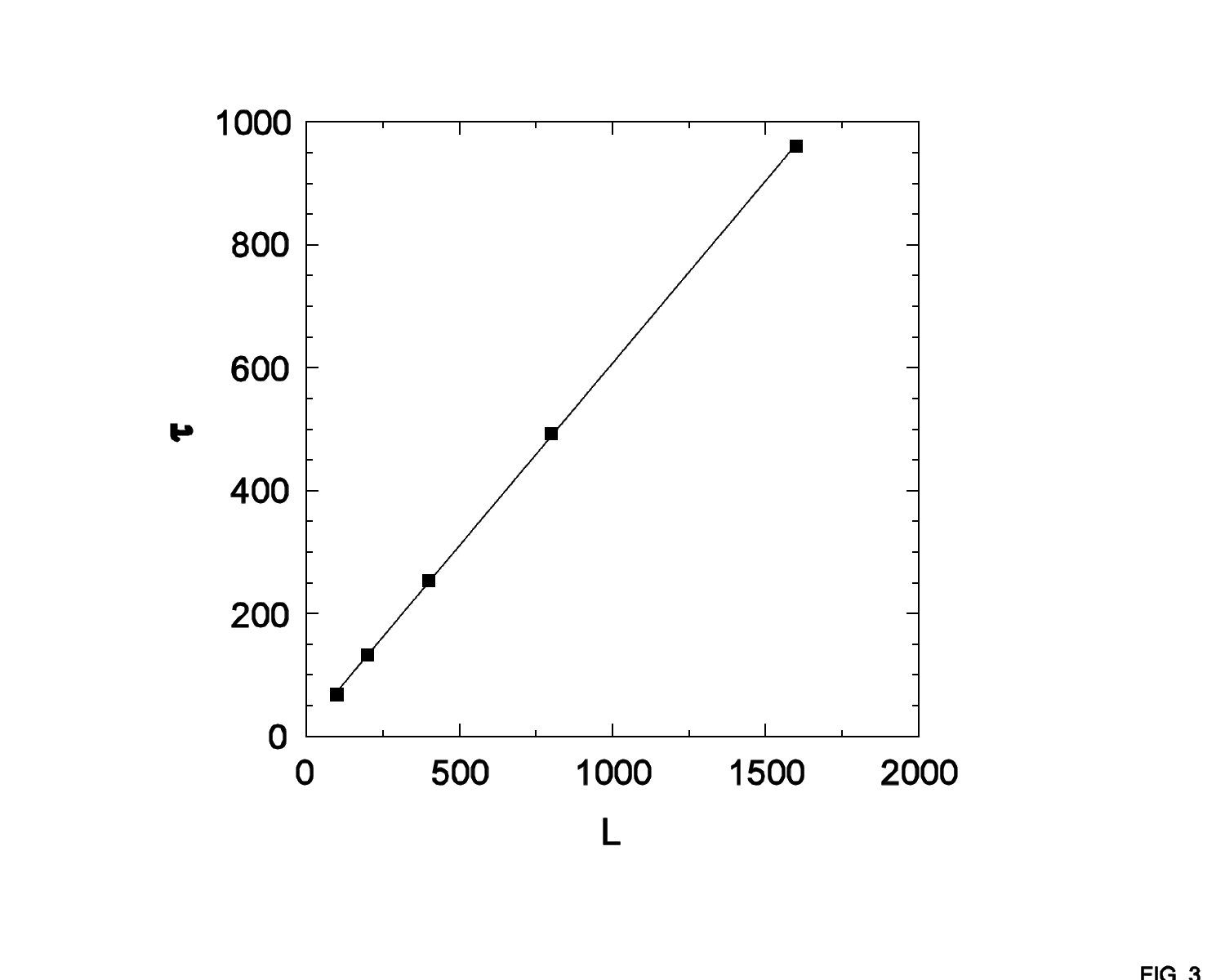}
\caption{%
Lifetime $\tau$ versus system size as in Figure~\ref{fig2}.
}
\label{fig3}
\end{figure}

Power-law scaling of $\rho_a$ and $\tau$ with $L$ provides an effective
criterion for locating the critical point in simulations; we use it to
determine $p_c$ for $\zeta = 0.25$, 0.5 and 0.75. Our results agree, to within
statistical uncertainty, with the mean-field prediction $p_c = 1-\zeta$. In all
three cases we find \[\beta/\nu_\perp = 0.5\] to within a statistical uncertainty
of less than 0.5\%, strongly suggesting the value 1/2 for this exponent ratio
(see Figure~\ref{fig2}). The lifetime $\tau$ (Figure~\ref{fig3}) can be fit to high precision with
the expression $\tau = C + c' L$, where $C$ and $c'$ are constants, so that the
ratio \[\nu_{||}/\nu_\perp = 1.\]  Finally, the fact that the order parameter
$\rho_a$ is proportional to $\zeta-\zeta_c$ near the transition implies the
exponent value \[\beta=1.\]  Using the finite-size scaling relations we then have
\[\nu_\perp = \nu_{||} = 2.\]

The moment ratio $m$ has been found to take a well-defined value at the critical point of an absorbing-state phase transition~\cite{dickman-kamphorstlealdasilva-98,kamphorstlealdasilva-dickman-99}.
Consistent with this result we find $m \to m_c = 1.298(4)$ at all three $\zeta$ values studied.
We note that while $\beta = 1$ is characteristic of mean-field-like transitions to an absorbing state, the values of $\nu_\perp$, $\nu_{||}$ and $m_c$, are not typical of other known universality classes for absorbing-state phase transitions~\cite{odor-04,hinrichsen-00,marro-dickman-99}.
We suspect that the anisotropic dynamics underlies this difference.

The model exhibits a somewhat different behavior at the end of the
critical line, $\zeta=1$, $p=0$.  In this case an isolated
particle goes to sleep at rate 1, i.e., it can never jump forward.
Simulations reveal no quasi-stationary state at this point: the
activity density decays to zero monotonically, for all system
sizes ($L=100$,...,800) investigated.  The activity density again
grows $\propto \zeta-1$ near the transition, so that $\beta$
retains its value of unity.

\subsection{Approach to the steady state}

We studied two kinds of initial condition. In one, $N=\zeta L$ active walkers
are inserted randomly and independently into the system; in the other (for
$\zeta=1/2$), only even-numbered sites are initially occupied by an active
walker. (We call these {\it random} and {\it alternating} initial
configurations, respectively. In studies with random initial conditions each
realization is performed using a different, independent initial configuration.)
The two initial states lead to the same quasi-stationary properties, but the
approach to the latter is different in the two cases.

\begin{figure}[!htb]
\centering
\includegraphics[width=10cm]{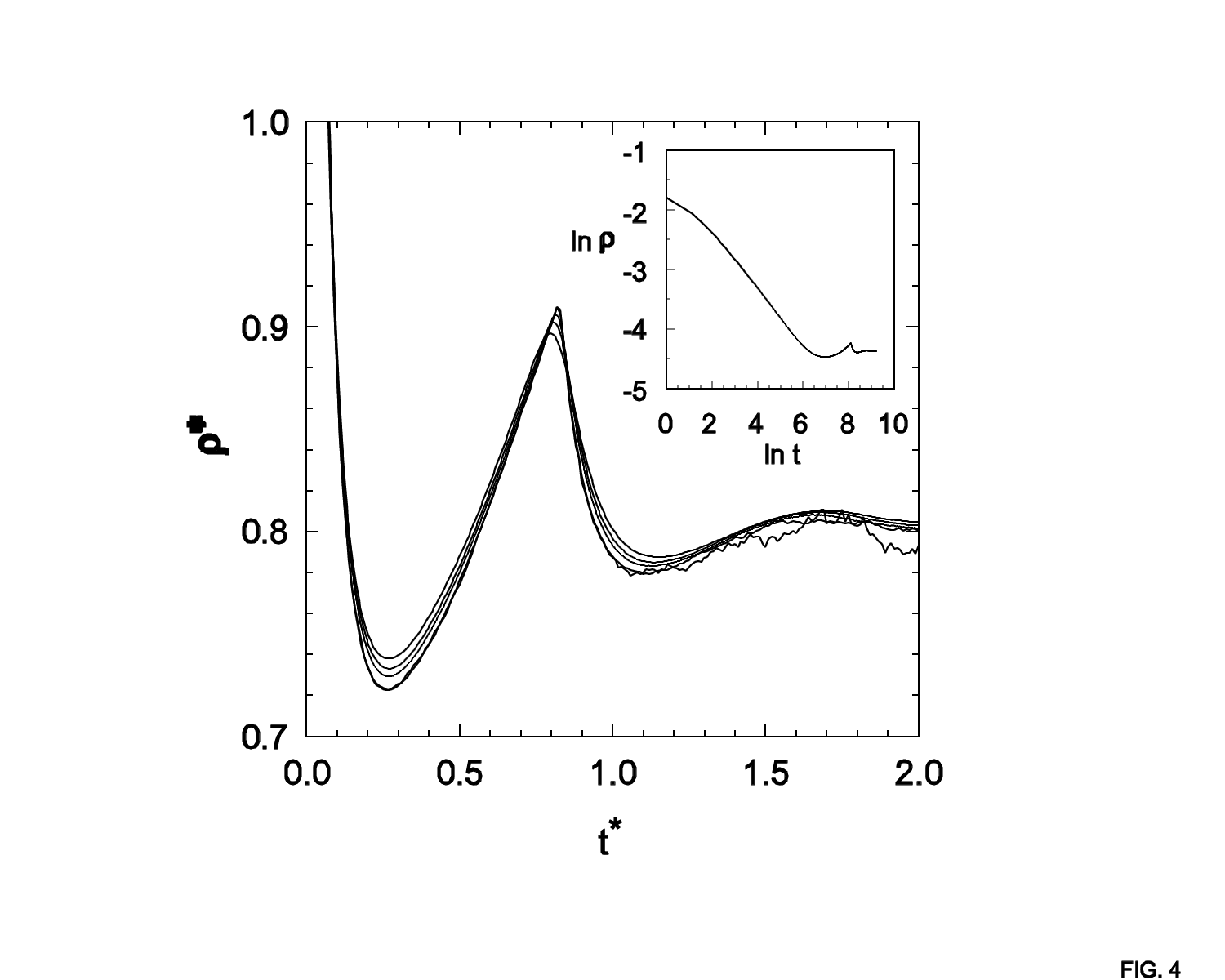}
\caption{%
Main graph: scaled activity density $\rho^* = L^{1/2} \rho_a(t)$ versus scaled time $t^* = t/L$, for $\zeta=p=1/2$ with an alternating initial configuration.
Data for system sizes $L=200$, 400, 800, 4000 and 8000
are superposed (sharper maximum corresponds to larger size). Inset: activity
density versus time on log scales, L=4000.
}
\label{fig4}
\end{figure}

Consider first the evolution of $\rho_a$ (averaged over $10^5$ independent
realizations) at the critical point $\zeta = p = 1/2$, using the alternating
initial configuration.
Figure~\ref{fig4} shows that the evolution is nonmonotonic, as has been found for the stochastic sandpile (with symmetric dynamics) at its critical point~\cite{dickman-03}.
The main graph of Figure~\ref{fig4} shows $\rho^* \equiv L^{1/2} \rho_a(t) $ as a function of $t^* = t/L$.
(The definitions of the
scaling variables $\rho^*$ and $t^*$ are motivated by the finite-size scaling
results discussed above.) Under this rescaling, data for $L=200$, 400, 800,
4000 and 8000 collapse onto a master curve.  (The collapse is not perfect; the
secondary maximum near $t^* = 0.8 $ becomes sharper as the system size is
increased.)

The inset of Figure~\ref{fig4} shows the overall relaxation to the quasi-stationary state.
The initial decay appears to follow a power law $\rho_a \sim t^{-\delta}$ with
$\delta = 0.50$.  At absorbing state phase transitions one expects the scaling
relation $\delta = \beta/\nu_{||}$, which is indeed verified if we insert the
values $\beta = 1$ and $\nu_{||} = 2$ found above.  The initial growth of the
moment ratio is expected to follow $m-1 \sim t^{1/z}$, with the dynamic
exponent $z$ equal to the ratio $\nu_{||}/\nu_\perp$~\cite{silva-dickman-drugowichdefelcio-04}.
In fact we find $m - 1 \propto t$, consistent with the exponent ratio found above.

For random initial conditions the general picture is similar, although there
are some differences in detail.  The relaxation is again nonmonotonic, with a
collapse of data for various system sizes using the scaling variables $\rho^*$
and $t^*$, but the secondary maximum (which falls near $t^* = 0.7$), is smooth,
rather than cusp-like as for the alternating initial condition. The initial
decay again appears to follow a power law, with $\delta \simeq 0.51$. The
initial growth in the moment ratio follows $m-1 \sim t^{1/z}$, but with $1/z =
1.025$ rather than the expected value of unity. The slightly larger apparent
exponents observed with random initial conditions may reflect corrections to
scaling due to relaxation of long-wavelength modes present in the initial
distribution (and which are strictly excluded in the alternating case).

\begin{figure}[!htb]
\centering
\includegraphics[width=10cm]{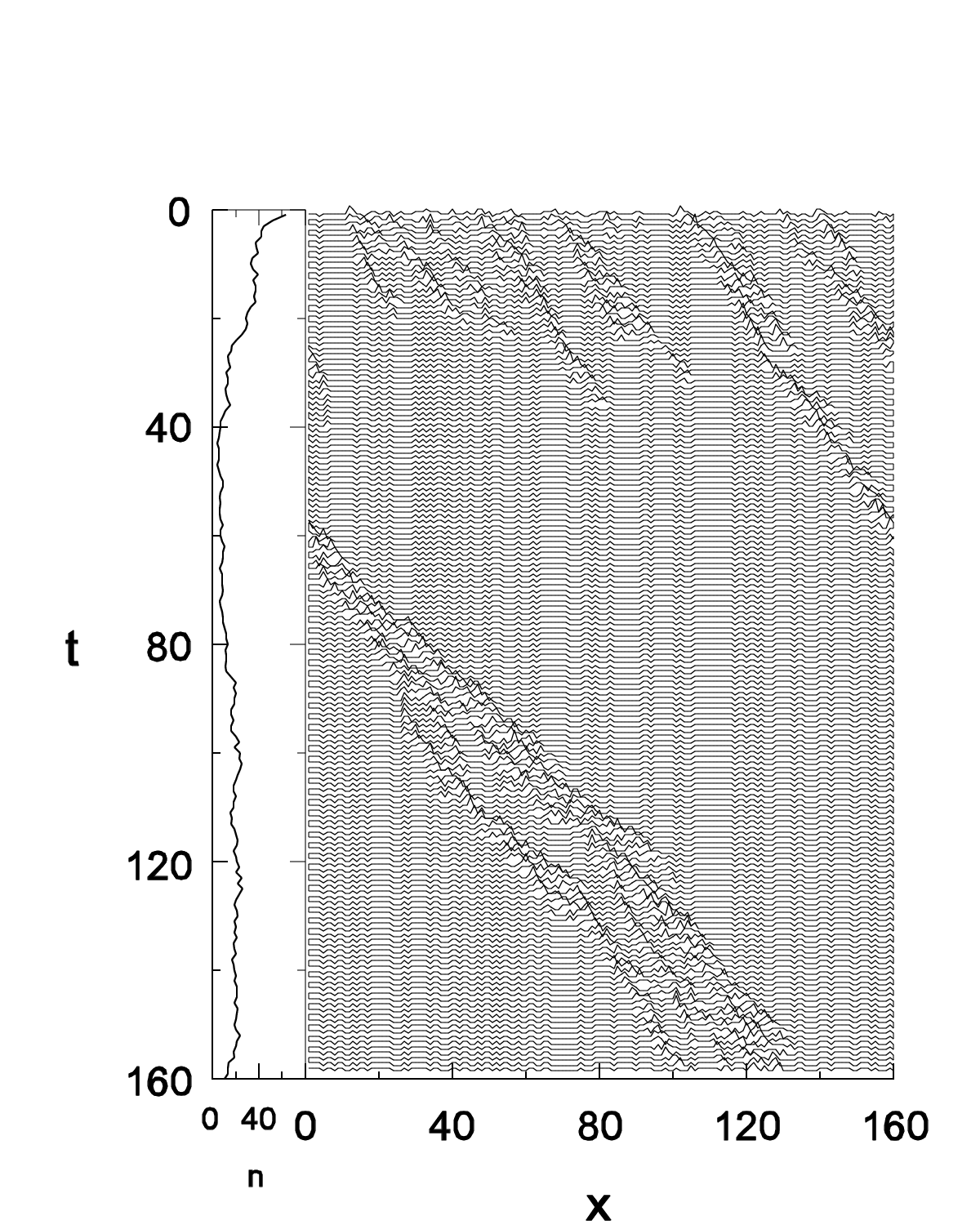}
\caption{%
Typical evolution of a system at the critical point $\zeta =
p = 1/2$, with a random initial configuration, $L=160$.  In each horizontal
sweep, the height of the line represents the number of particles at site $x$,
with a sleeping particle corresponding to height -1.
The graph at the left shows the number $n$ of active particles versus time.
}
\label{fig5}
\end{figure}

Some understanding of the relaxation may be gleaned from the spatial
distribution of the particles.  Figure~\ref{fig5} shows the spatio-temporal evolution of a
typical realization at the critical point, $\zeta = p = 1/2$, for random
initial conditions.  We see that after an initial transient, all of the active
particles are confined to a relatively narrow band.  During coalescence into a
single band, the activity density decreases rapidly; coalescence appears to be
irreversible. Studies of larger rings confirm these observations. The time for
the activity to become confined to a single band grows with system size, but is
typically smaller than $L/2$.  (We have not determined if the time grows
linearly with $L$ or more slowly.)  The active region propagates through the
system at a steady rate, expanding or contracting due to intrinsic fluctuations
and to the varying density of sleeping particles it encounters as it moves.  The
boundaries of the active region move at a speed somewhat greater than unity: in
large systems the speed is 1.18 - 1.20 (sites per unit time). Repeated
encounters of the active band with regions rich in sleeping particles may be
connected with the revivals observed in the activity density (Figure~\ref{fig4}). We
observe the coalescence into a single band of activity, and the same speed of
propagation, in studies with the alternating initial configuration. (Naturally
the initial transient is different in the two cases.) In the infinite system,
activity cannot be confined to a narrow band, but we should expect, on the
basis of the foregoing observations, a steady {\it coarsening} of the activity
pattern.

\begin{figure}[!htb]
\centering
\includegraphics[width=10cm]{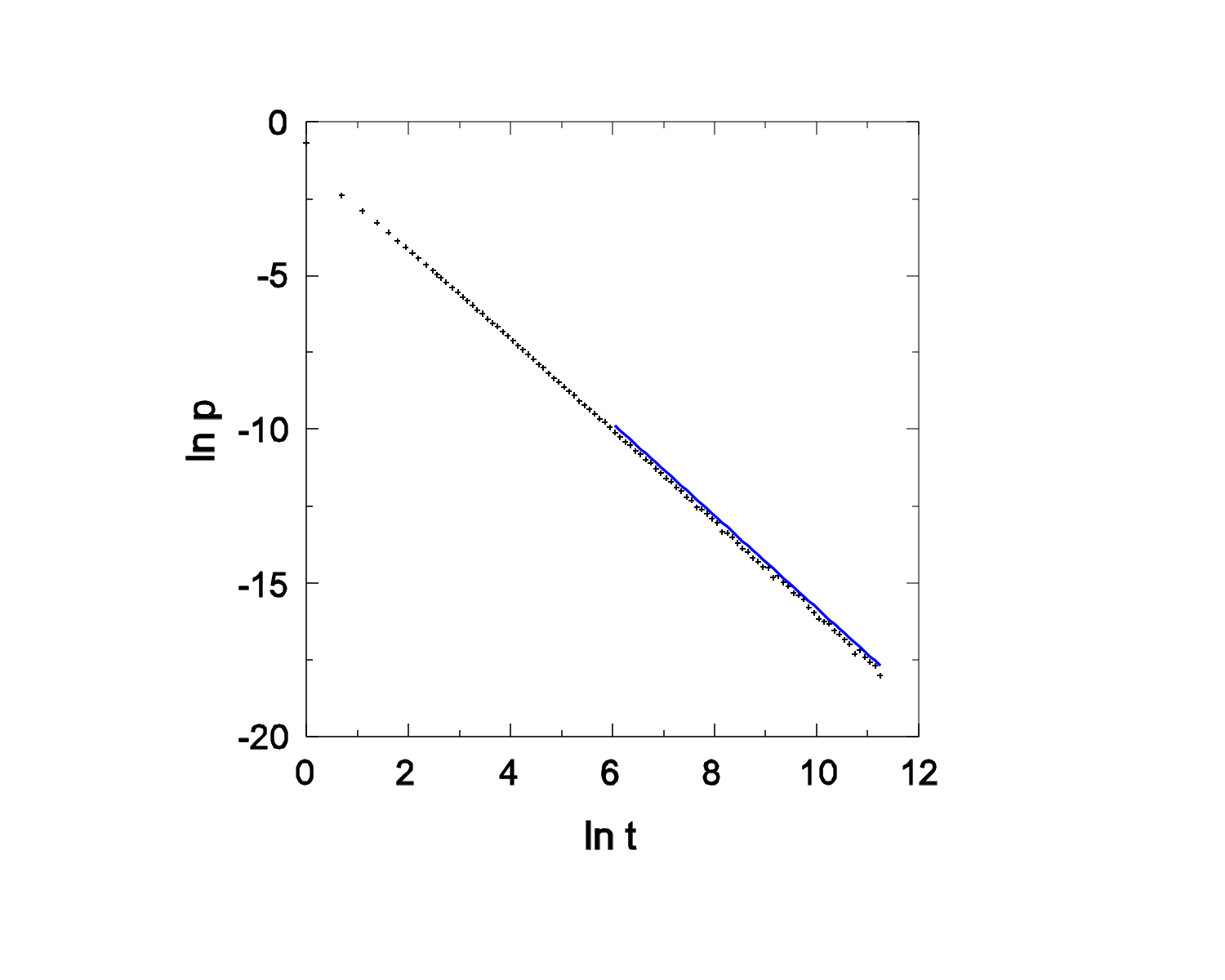}
\caption{%
Distribution density of the time of the first jump to the origin for a system at the critical point $\zeta = p = 1/2$, with an alternating initial configuration, and $L=10^5$ sites.
The blue line has slope $-1.50$.
}
\label{fig5a}
\end{figure}

We have also studied the distribution of first passage times $\tau_0$ to the
origin of the ARW2 model on the line (i.e., in a system without periodic
boundaries); $\tau_0$ is defined as the time at which a particle first jumps
from site -1 to the origin.  To study its distribution we simulate the system
on the lattice extending from $x=-L$ to $x=0$.  We determine the probability
density $p(\tau_0) $ up to a certain maximum time, by studying a series of
lattice sizes $L$, until $p(\tau_0)$ stabilizes.
Figure~\ref{fig5a} shows the density obtained for a lattice size of $10^5$ sites, for a system at the critical point, $\zeta = p = 1/2$.
The same result is obtained for $10^6$ sites, within the uncertainty.
The data can be fit with a power-law,
$p(\tau_0) \sim \tau_0 ^{-\alpha}$, with $\alpha = 1.50(1)$.
Thus the mean first passage time to the origin diverges at the critical point.
It is curious that an interacting particle system with totally asymmetric jumps shares the same exponent as that of a symmetric random walk.
For an unbiased random walk the first return is the smallest time when it is positive, whereas $\tau_0$ is determined by the smallest lattice interval $[-x,0]$ where the presence of initial particles exceeds the gaps of activity.
This subtle heuristics explains why the tail of $\tau_0$ decays with exponent $1/2$, though it is not so clear why its density is so smooth.


\subsection{The symmetric case}

We performed a series of studies of the symmetric ARW2 model at density $\zeta = 1/2$.
In contrast to the asymmetric case, the critical value of the hopping probability, $p_c$, is considerably higher than its mean-field value, $p_c = \min \{\zeta, 1\}$.
In terms of the ARW1, it means that
\[
  \rho_c > \frac\lambda{1+\lambda},
\]
and there are partial mathematical results in this direction~(Theorem~\ref{theo1slwa}).
Studies using rings of up to 12800 sites yield $p_c =
0.87835(2)$.  (Our criterion for criticality is that the stationary activity
density follow a power law, $\rho_a \sim L^{-\beta/\nu_\perp}$.)  These studies
yield the exponent ratio $\beta/\nu_\perp = 0.23(1)$, very different from the
result of 0.5 found in the asymmetric case.  A study of the growth of $m-1$ at
the critical point, in a system of 12800 sites, yields the dynamic exponent $z
= 1.51(1)$. The stationary value of the moment ratio $m$ at the critical point
is $m_c = 1.15(1)$.
The results for $\beta/\nu_\perp$, $z$ and $m_c$ are all quite far from the corresponding values in the asymmetric case.
They are, on the other hand, rather close to those found for a conserved stochastic sandpile~\cite{dickman-06}: $\beta/\nu_\perp = 0.217(6)$, $z=1.50(4)$ and $m_c = 1.14(1)$.

\begin{figure}[!htb]
\centering
\includegraphics[width=10cm]{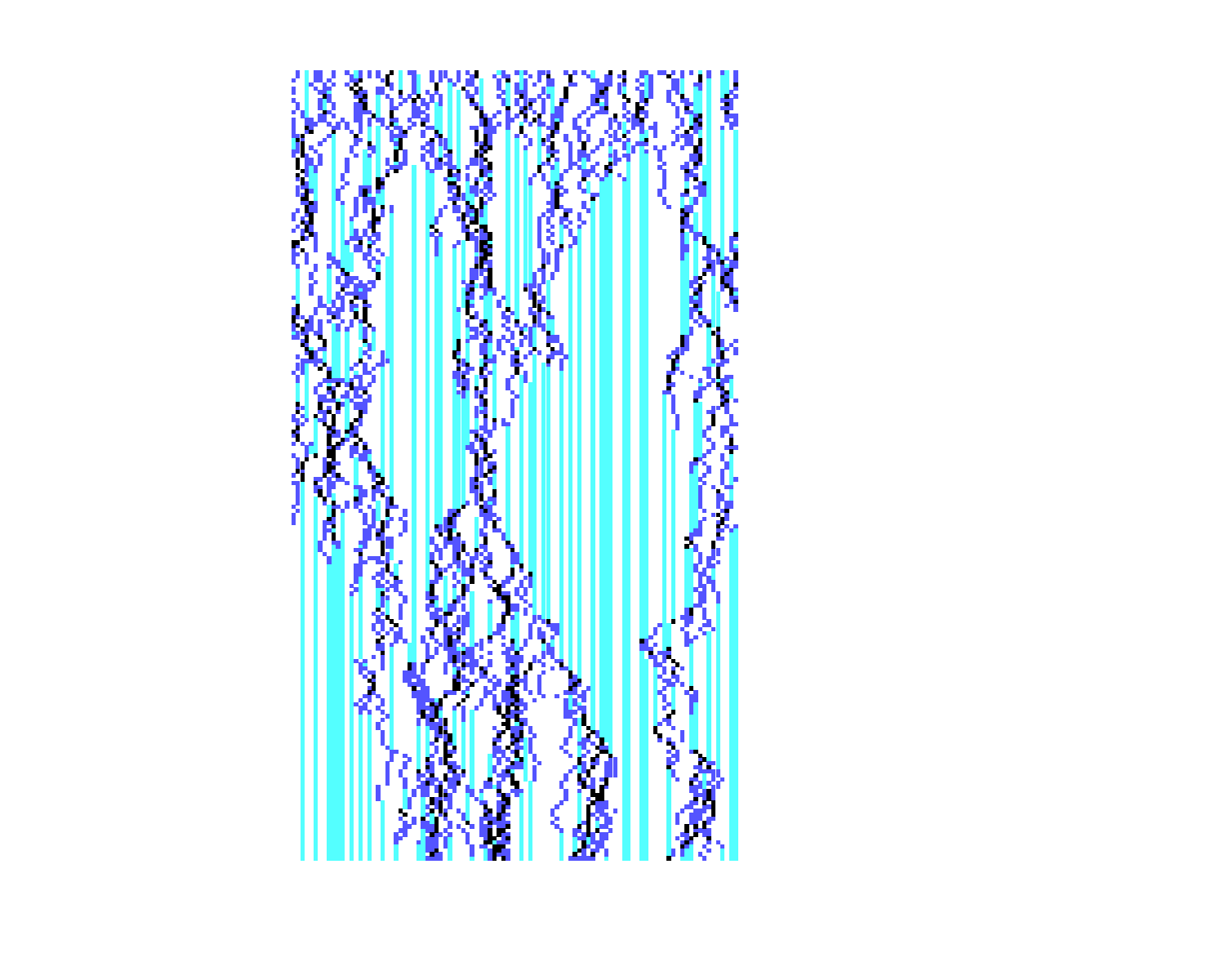}
\caption{
A typical realization of the symmetric ARW model near the critical point.
Time increases downward, with white, light blue, dark blue and black points
representing empty sites, sites with an inactive particle, sites with an active particle,
and sites with $\geq 2$ particles, respectively.
}
\label{fig6}
\end{figure}

These results support the assertion (based on considerations of symmetry) that
the symmetric ARW model falls in the conserved stochastic sandpile universality
class.  Since scaling properties of sandpile models are rather subtle, we defer
a full characterization of the symmetric model to future work.  It is
nevertheless clear that the symmetric and asymmetric ARW exhibit very different
critical behavior.  On a qualitative level the difference is quite dramatic if
we compare the evolution of the asymmetric model (Figure~\ref{fig5}) with that of the
symmetric model at its critical point (Figure~\ref{fig6}).
In the latter case there is no tendency for the activity to become confined irreversibly to a narrow band; active regions are seen to branch as well as coalesce.

As noted above, the model studied by Jain \cite{jain-05} corresponds to setting $p=0$ in the symmetric ARW.
The properties demonstrated in \cite{jain-05} are in fact very different than those obtained here, for $p > 0$.
In the former case the active phase, which exists for $\zeta > \zeta_c = 1$, has $\rho_a \propto \zeta - \zeta_c$, so that the critical exponent $\beta = 1$, and the stationary probability distribution in the active phase is uniform on the set of allowed configurations (i.e., those in which no site is empty).
Since the active phase has a product measure, correlation functions are identically zero and the critical exponent $\nu_\perp$ is undefined.
The difference between the scaling properties observed for $p > 0$, and those found in \cite{jain-05} may be understood by noting that in our case an isolated particle, while active, may evade becoming immobile by jumping to an occupied site.
In this process it may reactivate a sleeping particle.
Thus the number of active particles fluctuates while total particle number is conserved, a hallmark of models in the conserved stochastic sandpile universality class.
When $p=0$, by contrast, the number of mobile particles is \emph{fixed} at $N-L$ in the stationary state.

It is also worth noting that although the \emph{asymmetric} ARW (with $p>0$) shares the critical exponent value $\beta=1$ with the model studied in \cite{jain-05}, it is different since the exponent $\nu_\perp$ is well defined, and the model satisfies finite-size scaling.
Again, the asymmetric ARW features a fluctuating number of active particles when $p>0$.


\section{Rigorous results, conjectures and open problems}
\label{sec4resultschalenges}

In this section  we will summarize the few existing rigorous results concerning the ARW model, discuss some conjectures and present several open problems, whose understanding may shed some light on the long-time behavior of the system. 

A rigorous understanding of this model is still in its embryonic stage, and some of the open questions appear to be quite difficult and mathematically challenging.

\subsection{General case}
We start with the first basic fact, proved in~\cite{rolla-sidoravicius-09} using the Diaconis-Fulton~\cite{diaconis-fulton-91,eriksson-96} representation of the model.
The representation provides an Abelian property for the dynamics of the system with finitely many particles, and -- what is particularly important -- provides monotonicity for the occupation times in $\zeta$ as well as in $\lambda$.
\begin{theorem}
[\cite{rolla-sidoravicius-09}]
For $d\geqslant 1$ and any translation-invariant random walk and $\lambda>0$, there exists $\zeta_c \equiv \zeta_c (\lambda)\in[0,\infty]$, such that if the initial distribution is i.i.d\ Poisson with density $\zeta$ then
\[
 P(\mbox{system locally fixates}) =
\begin{cases}
 1, & \zeta<\zeta_c \\
 0, & \zeta>\zeta_c.
\end{cases}
\]
Moreover, $\zeta_c$ is non-decreasing in $\lambda$.
\end{theorem}

For fixed $\lambda$ the value of $\zeta_c (\lambda)$ is not known, however some theoretical arguments suggest, and numerical simulations support, that the following holds:
\begin{conjecture}
For any dimension, any random walk, and any $\lambda >0$,
\[ 0<\zeta_c (\lambda)<1. \]
\end{conjecture}

\subsubsection{Supercritical regime}

Using Peierls type argument one can show that $\zeta_c (\lambda)<+ \infty$:

\begin{theorem}
[\cite{kesten-sidoravicius-06}]
\label{theo1finite}
Consider simple symmetric random walks on $\mathbb Z^d,d\geqslant 1$.
There exists $\zeta_0<\infty$ such that $\zeta_c (\lambda)<\zeta_0$ for all $\lambda$.
\end{theorem}
Recently E. Shellef improved this estimate:
\begin{theorem}
[\cite{shellef-09}]
\label{theo1density1}
Under the same hypotheses,
\begin{equation}
\label{pe1}
\zeta_c (\lambda) \leqslant 1.
\end{equation}
\end{theorem}
Another approach to prove~\eqref{pe1} is to show mass conservation for this model:
\begin{theorem}
[\cite{amir-gurelgurevich-09}]
For i.i.d.\ initial conditions, simple symmetric random walks, if there is local fixation, then each particle jumps finitely many times.
By the mass transport principle, this implies that whenever there is fixation the density of the limiting state in the same as the density of the initial state, in particular it implies~\eqref{pe1}.
\end{theorem}

However, the following problem remains open:
\begin{problem}
Show strict inequality in~\eqref{pe1}.
\end{problem}

Another interesting question about the supercritical regime is the following:
\begin{problem}
\label{problem3}
For each $\zeta>\zeta_c$, show that there is a unique non-trivial invariant distribution, ergodic with respect to spatial translation, whose particle density is $\zeta$.
\end{problem}

\subsubsection{Subcritical behavior}

On the other hand, it is rather easy to get convinced that $\zeta_c$ ought to be strictly positive. Despite the recent progress in the one-dimensional case (see \cite{rolla-sidoravicius-09}), there are
no results in higher dimensions:
\begin{theorem}
[\cite{rolla-sidoravicius-09}]
\label{pe2}
For $d=1$, bounded range random walks and any $\lambda>0$ we have that $0<\zeta_c\leqslant 1$.
For nearest-neighbor walks we have:
\label{theo1slwa}
\[
 \frac \lambda{1+\lambda} \leqslant \zeta_c \leqslant 1.
\]
\end{theorem}
The proof of the above theorem again relies on the Diaconis-Fulton representation of the dynamics, in particular it uses the Abelian property and certain monotonicity for finite particle systems. Conceptually all the ingredients of the proof can be used in higher dimensions, with the exception of the last estimate, which in dimensions $\geqslant 2$, if repeated straight forwardly, boil down to the need for refined bounds on growth intensity of a Diffusion Limited Aggregation type growth model, and the analog
of the one-dimensional argument produces unsatisfactory estimates.
Thus, we have
\begin{problem}
Show that $\zeta_c>0$ for $d\geqslant 2$.
\end{problem}

Though Theorem~\ref{pe2} establishes the fact that $\zeta_c>0$ in one-dimension for a rather broad class of walks, it does not
give a satisfactory description of the final absorbing state, nor correct estimates for the fixation time. We may therefore state several
important questions.

\begin{problem}
Describe the distribution of the final configuration after fixation (in any dimension, including $d=1$).
Numerical analysis suggests that it depends on the initial distribution.
It seems however to be less sensitive when $\zeta\sim\zeta_c$, when the final state is prominently different from a Bernoulli.
\end{problem}

\begin{problem}
Establish the rate of fixation, i.e., the asymptotic behavior of the probability that there is an active particle at the origin after time $t$, for $t$ large enough. We believe that if $\zeta $ is small enough, the decay should be exponential. However when $\zeta $ approaches $\zeta_c $ we do not exclude the possibility that decay may become slower (stretched exponential, or even algebraic).
\end{problem}

The following problems constitute possible intermediate steps to understand the behavior of the system in the subcritical regime.

\begin{problem}
Consider an infinite volume system with only $k < + \infty$ particles, which are all initially located at the origin and in state $A$.
Let $\widehat\tau_k$ denote the (a.s.\ finite) time when the system fixates, i.e., when all particles become inactive.
For any $k >0$ find an asymptotic bound for $P(\widehat\tau_k > n)$ as $n \to + \infty$.
Prove the Large Deviation Principle for $\widehat\tau_k$.
\end{problem}

\begin{problem}
As before, consider a finite system of $k$ particles
and denote by $L_k (s)$ and $R_k (s)$ positions of the leftmost and rightmost
particles in the system, and by $L^A_k (s)$ and $R^A_k (s)$ position of the
leftmost and rightmost $A$-particles in the system -- which do not necessarily
coincide with $L_k (s)$ and $R_k (s)$, but for all times satisfy inequalities
$L^A_k (s) \ge L_k (s)$ and $R^A_k (s) \le R_k (s)$.

How does $R^A_k (s) - L^A_k (s)$ behave during the time
interval $[0, \widehat\tau_k ]$?

What can we say about the distribution of $R_k
(\widehat\tau_k) - L_k (\widehat\tau_k)$, the diameter of the configuration in
the final state, and about its displacement with respect to the origin $|0 -
L_k (\widehat\tau_k)|$?
\end{problem}

\subsubsection{At criticality}

We have good reason to believe that at the critical density the system does not fixate in any dimension (see comments in the next subsection). 
However time intervals between successive visits of active particles to the origin will diverge to infinity. It is not clear that here we
are observing an aging phenomenon in one of accepted senses.

\subsection{Totally asymmetric dynamics in one dimension}

The description of the totally asymmetric walk in one dimension is relatively well understood. 
Let $x_0(t)$ denote the position of a tagged particle starting at the origin.
\begin{theorem}
[\cite{hoffman-sidoravicius-04}]
\label{theo1tarw}

For $d=1$, and the totally asymmetric walk,
\[
 \zeta_c = \frac \lambda{1+\lambda}.
\]

If $\zeta < \zeta_c$, then
\[
P[\eta^A_s(0) > 0 \mbox{ for some } s \geq t] \leqslant
 c_1 e^{-c_2 t}.
\]

If $\zeta > \zeta_c$,
then
 \begin{equation}
  \lim_{t\to \infty} \frac {x_0 (t)}t = v_d > 0.
 \label{current}
 \end{equation}

If $\zeta = \zeta_c$, and initially all particles in the system are $A$-particles, then the system does not fixate (!).
However,
 \begin{equation}
 \label{eq5nocurrent}
 \lim_{t\to \infty} \frac {x_0 (t)}t = 0.
 \end{equation}
\end{theorem}

The limit~(\ref{current}) implies that there is a current and that the tagged particle has an asymptotic velocity.
This is related to Problem~\ref{problem3}, as the limiting speed should be the density of active particles in the (unique) non-trivial ergodic invariant distribution with total density of particles $\zeta$.

The limit~\eqref{eq5nocurrent} tells us that there is no current in the system.
This should be related to the absence of a non-trivial ergodic invariant distribution with total density $\zeta_c$, i.e., the phase transition is not of first order.

Moreover, it motivates the following problem.
Consider a system with critical density, and denote by $\widetilde \tau_i (0)$ the time spent between arrivals of the $(i-1)$-th and $i$-th $A$-particle at the origin.
\begin{problem}
{Characterize the behavior of $\widetilde \tau_i (0)$ when $i \to + \infty$.}
\end{problem}
We believe that the system exhibits behavior reminiscent of so-called aging for disordered systems.

\begin{problem}
Explain the origin of the `spikes' in the time-dependent density of active particles in Figure~\ref{fig4}).
\end{problem}

\section*{Acknowledgments}

We are grateful to Chris Hoffman for fruitful discussions and to Harry Kesten and Jeff Steif for many valuable comments.
L.~T.~Rolla thanks the hospitality of CWI, where part of this research was done.
This work was supported by CNPq grant 141114/2004-5, FAPERJ, FAPESP grant 07/58470-1, and FSM-Paris.
R.~Dickman acknowledges support from CNPq.

\nocite{benhough-krishnapur-peres-virag-06}
\nocite{karlin-mcgregor-59}

\baselineskip 12pt
\frenchspacing
\bibliographystyle{bib/siamrolla}
\bibliography{bib/leo}

\end{document}